\documentclass[%
11pt,
reprint,
onecolumn,
tightenlines,
superscriptaddress,
preprintnumbers,
nofootinbib,
amsmath,amssymb,amsthm,
physrev,
eqsecnum,tikz,
]{revtex4-2}
\usepackage{isomath}
\usepackage{amsmath,amsthm}
\usepackage{amsbsy}
\usepackage{amssymb}
\usepackage{amscd}
\usepackage{amsfonts}
\usepackage{stmaryrd}
\usepackage{siunitx}
\usepackage{euscript}
\usepackage[utf8]{inputenc}
\usepackage[T1]{fontenc}
\usepackage{newtxtext} 
\everymath{\displaystyle}
\usepackage{exscale}
\usepackage{microtype}
\usepackage{booktabs}
\usepackage{algorithm}
\usepackage{algpseudocode}
\usepackage{hyperref}

\usepackage{graphicx}
\usepackage{epstopdf}
\usepackage{boxedminipage}
\usepackage{calc}
\usepackage[dvipsnames]{xcolor}
\graphicspath{ {media/} {media/} }
\usepackage[caption=false,justification=raggedright]{subfig}

\usepackage{setspace}
\usepackage{enumitem}
\setitemize{noitemsep,topsep=0pt,parsep=0pt,partopsep=0pt}
\setenumerate{noitemsep,topsep=0pt,parsep=0pt,partopsep=0pt}
\setdescription{noitemsep,topsep=0pt,parsep=0pt,partopsep=0pt}

\usepackage[normalem]{ulem}

\usepackage{orcidlink}
\usepackage{siunitx}
\usepackage[small]{titlesec}

\titlespacing*{\section}{0pt}{12pt plus 4pt minus 2pt}{2pt plus 2pt minus 2pt}
\titlespacing*{\subsection}{0pt}{12pt plus 4pt minus 2pt}{2pt plus 2pt minus 2pt}
\titlespacing*{\subsubsection}{0pt}{12pt plus 4pt minus 2pt}{2pt plus 2pt minus 2pt}
\titlespacing*{\paragraph}{0pt}{12pt plus 4pt minus 2pt}{2pt plus 2pt minus 2pt}

\makeatletter

    \renewcommand*{\p@subsection}{}
    
    \renewcommand*{\p@subsubsection}{}
\makeatother

\usepackage{isomath}
\usepackage{amsmath}
\usepackage{amssymb}
\usepackage{amscd}
\usepackage{amsfonts}
\usepackage{upgreek}

\theoremstyle{definition}

\newcommand{\bfsigma}{\mathbold {\sigma}}

\newcommand{\bfpsi}{\boldsymbol{\psi}}

\DeclareMathOperator{\divergence}{div}

\DeclareMathOperator{\variation}{\updelta}

\newcommand{\dm}{\ \mathrm{d}}
\newcommand{\deriv}[2]{\frac{\dm #1}{\dm #2}}

\newcommand{\bfn}{{\mathbold n}}

\newcommand{\bft}{{\mathbold t}}
\newcommand{\bfu}{{\mathbold u}}

\newcommand{\bfx}{{\mathbold x}}

\newcommand{\bfC}{{\mathbold C}}

\usepackage{subfig}
\usepackage{graphicx}
\usepackage{makecell}

\usepackage{array}
\newcolumntype{L}{>{$}l<{$}}
\newcolumntype{R}{l}

\begin{document}


\preprint{To appear in Journal of Applied Mechanics (DOI: \href{https://doi.org/10.1115/1.4071127}{10.1115/1.4071127})}

\title{A Kinetic Phase-Field Model of Diffusion Bonding: A Nonlocal Approach to Interface Coalescence}

\author{Maryam Khodadad}
    \email{mkhodada@andrew.cmu.edu}
    \affiliation{Department of Civil and Environmental Engineering, Carnegie Mellon University}

\author{Noel Walkington}
    \affiliation{Center for Nonlinear Analysis, Department of Mathematical Sciences, Carnegie Mellon University}

\author{Suresh Kalyanam}
    \affiliation{Westinghouse Electric Company LLC}

\author{Matteo Pozzi}
    \affiliation{Department of Civil and Environmental Engineering, Carnegie Mellon University}

\author{Kaushik Dayal}
    \affiliation{Department of Civil and Environmental Engineering, Carnegie Mellon University}
    \affiliation{Center for Nonlinear Analysis, Department of Mathematical Sciences, Carnegie Mellon University}
    \affiliation{Department of Mechanical Engineering, Carnegie Mellon University}

\date{\today}


\begin{abstract}
    Conventional phase-field models often drive solid-solid interfaces to coalesce when in close proximity. 
    This feature limits their use for processes like diffusion bonding, where the interfaces might need to remain distinct under certain thermodynamic conditions. 
    We develop a kinetic phase-field model to address this problem, using an evolution equation based on a geometric conservation law for interfaces, rather than the gradient descent evolution that is typical in phase-field modeling.
    This formulation enables us to specify complex kinetic laws, and we use this to incorporate a physically-motivated geometric criterion to control interface merging. 
    This criterion, based on nonlocal higher-derivative curvature invariants of the phase field, can be temperature-dependent, allows for a range of behaviors from complete coalescence to the preservation of distinct boundaries. 
    Simulations show controlled bonding kinetics, demonstrating capabilities that are not available with existing methods for modeling interfaces that must remain distinct under given thermodynamic conditions.
\end{abstract}

\maketitle

\section{Introduction}

Solid-state diffusion bonding is an important joining technology in advanced manufacturing where interface structure plays a decisive role in system performance and safety \cite{li_two_2023, li_advances_2025, ramlow_joining_2025, mir_recent_2021}.
It is used in nuclear power applications \cite{zhang_enhanced_2023, peng_development_2024, li_alloy_2008}, energy storage systems \cite{xiao_understanding_2020, yang_interfacial_2021, de_prado_solid-state_2024}, where electrode-electrolyte interfaces require precisely controlled diffusion characteristics, and in aerospace components \cite{wu_effect_2018, kumar_vacuum_2022}, where dissimilar material joints must withstand extreme thermal cycling.  
Unlike conventional fusion-based joining methods, diffusion bonding creates high-integrity metallurgical connections without inducing melting or macroscopic deformation of the base materials~\cite{zhang_influence_2025, bajpai_spark_2022, li_microstructural_2022}. 
This characteristic is essential as the resulting interface microstructure critically influences the overall performance and reliability in demanding applications.

Ceramic and refractory materials present significant challenges in diffusion bonding processes due to their thermodynamic stability and complex interfacial behaviors. Recent experimental studies on ZrC-SiC composites by Lin et al.~\cite{lin_rapidly_2023} have demonstrated successful ceramic-ceramic interfaces using titanium interlayers via pulsed electric current joining, while He et al.~\cite{he_microstructure_2024} advanced the understanding of carbon-based composite brazing. The interfacial phenomena during ceramic diffusion bonding involve multiple physical mechanisms operating simultaneously, including solid-state diffusion, elastic deformation, phase transformation, and interfacial energy minimization ~\cite{ramlow_joining_2025}. These mechanisms create a multi-physics problem that requires sophisticated modeling approaches to capture the resulting microstructural evolution accurately.

\paragraph*{Prior work.}

The complex, multi-physics nature of interfacial phenomena in diffusion bonding has been studied using approaches ranging from continuum models \cite{bonet_numerical_1990, wang_stress-induced_2009, wu_dynamic_2007} to atomistic simulations \cite{wang_experimental_2024, ling_diffusion_2024, shi_interfacial_2016}.
Here, phase-field modeling offers a useful framework to capture the dynamics of interfaces across multiple scales. Its ability to represent diffuse interfaces and naturally handle complex morphological and topological changes without explicit interface tracking makes it well-suited for many material science problems \cite{steinbach_phase-field_2009, zhao_development_2023, hakimzadeh_phase-field_2022,hakimzadeh_phase-field_2025,hakimzadeh2025crack}.

However, the application of phase-field methods to solid-state diffusion bonding is a relatively recent endeavor \cite{kovacevic_interfacial_2020, wang_phase-field_2024, hosseinabadi_diffuse_2022}, distinct from its more established use in solidification or additive manufacturing processes \cite{boettinger_phase-field_2002, chen_phase-field_2002, keller_application_2017, lu_phase_2018, yang_phase-field_2021-1, yang_phase-field_2021}. 
Early work by Steinbach et al.~\cite{steinbach_phase-field_2012, zhang_phase-field_2012} addressed strong non-equilibrium conditions at interfaces, including chemical potential jumps. Their phase-field approach utilized distinct concentration fields for each of the coexisting phases.
Instead of enforcing equilibrium partitioning (i.e., balance of chemical potentials across the interface), these phase-specific concentrations were coupled through kinetic equations governing inter-phase solute exchange. 
This framework enabled the modeling of finite interface dissipation and was demonstrated for phenomena such as solute trapping during rapid solidification.

Building on these ideas, particularly for diffusion bonding, Kovacevic et al. \cite{kovacevic_interfacial_2020} developed a phase-field model for ZrC-Ti systems where concentration-dependent interfacial energy, driven by carbon concentration jumps, was identified as the key thermodynamic force governing interlayer homogenization and critical thickness. 
Similarly, Hosseinabadi \cite{hosseinabadi_diffuse_2022} modeled alumina laminate bonding, emphasizing the role of interfacial chemical reactions and phase transitions driven by solute concentration at the interface. In contrast to these explicitly concentration-driven models, Wang et al. \cite{wang_phase-field_2024} approached the diffusion bonding of 316H stainless steel by employing a multi-phase-field grain growth model. Their work focused on simulating microstructural evolution, including grain nucleation and growth across the bonding line, under various temperature profiles and mechanical loads, where temperature effects were primarily incorporated through a temperature-dependent grain boundary mobility.

\paragraph*{Contributions of this paper.}
Despite significant advancements, a core challenge persists in applying phase-field models to solid-state diffusion bonding: the inherent tendency of conventional evolution equations (e.g., Allen-Cahn type, where the time-derivative of the order-parameter is proportional to the variational derivative of the free energy) to promote interface coalescence. Driven by free energy minimization, and including gradient energy terms, these models can cause the gradual diffusion and disappearance of unbonded interfaces or thin interlayers over time, even when their persistence is physically expected. While some approaches, such as the concentration-dependent interfacial energies in Kovacevic et al.~\cite{kovacevic_interfacial_2020}, offer control by creating local energy minima, the phase-field dynamics may favor interface broadening or coalescence if these delicate energy balances are not perfectly maintained or if kinetic parameters are not numerically tuned. True interface preservation or arrested coalescence often relies on such precise energetic balancing or numerical adjustments, rather than an explicit mechanism for interface coalescence in the diffusion bonding process. This identifies a need for phase-field frameworks that intrinsically incorporate conditions for stopping interface motion based on alternative physical or geometric considerations.

Our work addresses this gap by directly introducing this physical effect into the phase-field kinetic law.
Our approach uses a kinetic phase-field model \cite{agrawal_dynamic_2015-1,agrawal_dynamic_2015} that is based on a geometric conservation law for interfaces rather than the typical phase-field gradient descent.
This provides the ability to specify arbitrary complex kinetic laws; in this work, the kinetic equation is designed to activate based on specific, nonlocal geometric criteria of the phase field itself---such as curvature and gradient invariants---that signify a stable, unbonded interface or a persistent thin layer. By modulating the effective interface mobility,  it provides a direct mechanism to suppress or arrest phase evolution locally, offering a physically motivated means to control interface coalescence rather than relying purely on complex energy landscapes or tuned mobility parameters. 
This geometric approach allows for a more flexible prediction of microstructure evolution, particularly in scenarios where experimental evidence suggests an unbonded interface or a distinct interlayer is the thermodynamically favored or kinetically arrested outcome.
This paper presents the detailed formulation and validation of this geometrically-controlled nonlocal phase-field model.

\paragraph*{Structure of the paper.}
 Section~\ref{sec:mathformulation} establishes the model framework, detailing the kinetic functional in one, two, and three dimensions, and discusses the thermodynamic consistency of the modified evolution equation. The finite element implementation scheme for the model is also described in this section. 
 Section~\ref{sec:numericalresults} then demonstrates the model capabilities through numerical simulations of model scenarios, illustrating the ability to control interface merging kinetics and predict diverse microstructural outcomes. 
 Section~\ref{sec:mechanical} discusses the coupling of this phase evolution with the mechanical response of the system, with a particular application to the ZrC-Ti diffusion bonding system previously studied by Kovacevic et al \cite{kovacevic_interfacial_2020}.

\paragraph*{Notation and Definitions.}

Boldface denotes vectors and tensor quantities. 
In 1D, partial derivatives with respect to the spatial coordinate $x$ are denoted by a subscript, e.g., $\phi_x := \partial\phi/\partial x$, and $\phi_{xx} := \partial^2\phi/\partial x^2$. 
In 2D and 3D, the operator $\nabla$ denotes the spatial gradient; the Hessian tensor is the gradient of the gradient, $\mathcal{H} := \nabla\nabla$; and $\nabla^2$ denotes the Laplacian.


\section{Kinetic Phase-Field Model Formulation}
\label{sec:mathformulation}

We develop a phase-field model that enables precise control over interface coalescence during diffusion bonding. Our approach introduces physically motivated constraints on interface merging, based on the geometric detection of interfaces, thereby addressing a fundamental limitation in conventional phase-field models where interfaces always merge when in proximity regardless of kinetic favorability.

\subsection{Phase-Field Evolution Equation}

The evolution of the phase field variable $\phi(\bfx,t)$, which implicitly describes the interface between distinct regions, is typically assumed to be governed by the Allen-Cahn equation:
\begin{equation}
\label{eq:allen_cahn}
    \frac{\partial\phi}{\partial t} = -M_{\phi}\variation_\phi \mathcal{P}
\end{equation}
where $M_{\phi}$ is a mobility coefficient;  $\mathcal{P}$ is the total free energy functional \cite{allen_microscopic_1979}; and $\variation_\phi \mathcal{P}$ is the functional derivative of $\mathcal{P}$ with respect to $\phi$.
This framework is utilized in models such as that by Kovacevic et al. \cite{kovacevic_interfacial_2020} for diffusion bonding, where the thermodynamic driving force $\mathcal{F} := -\variation_\phi \mathcal{P}$ incorporates chemical, elastic, and gradient energy contributions.

Although the Allen-Cahn equation (\ref{eq:allen_cahn}) is effective for many phenomena, there are challenges in prescribing complex interface kinetics and, critically, the nucleation of new phases. 
This makes calibration challenging and can lead to unphysical scenarios, such as nucleation occurring in a uniform phase driven solely by a large kinetic driving force, without meeting specific nucleation criteria \cite{abeyaratne2006evolution,dayal_kinetics_2006,chua2024interplay}.
To overcome these limitations and enable a clear separation between interface motion (kinetics) and interface creation (nucleation), Agrawal and Dayal \cite{agrawal_dynamic_2015-1, agrawal_dynamic_2015} proposed an alternative evolution law for $\phi$ derived from a geometric conservation principle for interfaces \cite{agrawal_dynamic_2015-1}. In their model, $\nabla\phi$ is interpreted as an interface density, and the resulting evolution equation is:
\begin{equation}
\label{eq:agrawal_evolution}
    \frac{\partial\phi}{\partial t} = |\nabla\phi| v_n^\phi + G
\end{equation}
Here, $v_n^\phi$ is a prescribed interface velocity governing the kinetics of existing interfaces, and can depend on various physical quantities such as stress or the thermodynamic driving force. 
$G$ is a source term that controls the nucleation of new interfaces, designed to activate only under specific, predefined critical conditions.

An important difference in the evolution law in \eqref{eq:agrawal_evolution} compared to \eqref{eq:allen_cahn} is the presence of $|\nabla\phi|$ in the kinetic response. 
This term ensures that kinetic evolution is active only at existing interfaces (where $|\nabla\phi| \neq 0$), effectively preventing the kinetic term from causing nucleation in uniform phase regions (where $|\nabla\phi| = 0$). Nucleation is then exclusively controlled by the separate term $G$.

The interface velocity $v_n^\phi$ in ~\eqref{eq:agrawal_evolution} encapsulates the kinetic response of the interface to thermodynamic driving forces. Generally, $v_n^\phi$ can be an arbitrary function of the driving force $\mathcal{F}$, subject only to the constraint of the second law of thermodynamics. For instance, a common assumption is linear kinetics, where $v_n^\phi = \kappa \mathcal{F}$, with $\kappa$ being a kinetic coefficient.

In our proposed model, we adopt the structure of (\ref{eq:agrawal_evolution}) but introduce a specific structure for the interface velocity $v_n^\phi$. 
We propose that the kinetic response to the thermodynamic driving force is modulated by a coalescence kinetic function, $g(\phi, \nabla\phi, \dots)$, which accounts for the nonlocal geometric state of the $\phi$ field, particularly to detect conditions indicative of interface proximity with thermodynamic thresholds for merging. 
The value of $g$ can range from $0$ (representing no hindrance to interface motion) to $1$ (representing complete arrest of the interface). The interface velocity is then expressed as:
\begin{equation}
    \label{eq:our_vn_phi_hindered}
    v_n^\phi (\mathcal{F},g) = \kappa (1-g) \mathcal{F}
\end{equation}
Substituting this definition into the kinetic part of ~\eqref{eq:agrawal_evolution}, and temporarily omitting the nucleation term $G$ to focus on kinetics, yields our central evolution equation for interface motion:
\begin{equation}
    \label{eq:our_evolution_hindered}
    \frac{\partial\phi}{\partial t} = -\kappa \left[1-g\left(\phi, \nabla\phi, \dots)\right ]\right |\nabla\phi| \variation_\phi \mathcal{P}
\end{equation}
Here, $\mathcal{P}$ represents a Ginzburg-Landau-type free energy, typically:
\begin{equation}
    \label{eq:free_energy}
     \mathcal{P}[\phi] = \int_\Omega \left( W(\phi) + \frac{1}{2}\kappa_{\phi} |\nabla\phi|^2 \right) \dm V  
\end{equation}
where $W(\phi)$ is the chemical free energy density (e.g., a double-well potential establishing distinct phases) and $\kappa_{\phi}$ is the gradient energy coefficient for the phase field order parameter. 
The variational derivative $\variation_\phi \mathcal{P} = \partial W/\partial\phi - \kappa_{\phi} \nabla^2\phi$ is the standard local driving force.

In this formulation (\ref{eq:our_evolution_hindered}), the term $\kappa (1-g)$ can be interpreted as an effective, spatially and temporally varying kinetic parameter. 
Crucially, unlike conventional kinetic coefficients or mobilities that are often treated as constants or simple functions of temperature, the factor $(1-g)$ depends on the evolving geometry of the phase field. 
This direct coupling between the geometric configuration and the kinetic response allows the system to locally modulate or even arrest interface motion when the conditions, as interpreted by $g$, deem coalescence unfavorable. 
This offers a distinct mechanism for controlling interface homogenization compared to models like Kovacevic et al. \cite{kovacevic_interfacial_2020}. 
While Kovacevic et al. employ a standard Allen-Cahn evolution (\ref{eq:allen_cahn}) with a numerically tuned constant mobility $M_{\phi}$, and achieve control through the coupling of $\phi$ to concentration fields and their influence on the free energy, our approach embeds the control  into the kinetic law via the geometric functional $g$.

\subsection{Geometric Control of Field Evolution in 1D}

The evolution of the phase field $\phi(x,t)$ in one spatial dimension, based on what was previously discussed (\ref{eq:our_evolution_hindered}) and incorporating the function $g$, can be expressed as:
\begin{equation}
    \label{eq:1d_hindered_evolution_general}
    \frac{\partial\phi}{\partial t} = -\kappa (1-g) |\phi_x| \variation_\phi \mathcal{P}
\end{equation}
where $\kappa$ is the kinetic coefficient. The free energy functional in one dimension is typically of the form:
\begin{equation}
    \mathcal{P}[\phi] = \int_L \left( \phi^2\left(1-\phi\right)^2 +\frac{\kappa_\phi}{2}|\phi_x|^2 \right) \dm x
\end{equation}
giving:
\begin{equation}
    \label{eq:1d_hindered_evolution_specific}
    \frac{\partial\phi}{\partial t} = \kappa (1-g) |\phi_x| \left( \kappa_\phi\phi_{xx} - 2 \phi(2 \phi-1)(\phi-1) \right)
\end{equation}
The term $|\phi_x|$ ensures that the kinetic evolution is localized to regions with varying $\phi$, i.e. interfaces, consistent with previous work \cite{agrawal_dynamic_2015}.

\subsubsection{Structure of the Coalescence Kinetic Function} 

The phase field $\phi(x,t)$ maps from a spatial-temporal domain, $(x,t) \in \Omega \times [0,T] \subset \mathbb{R} \times \mathbb{R}_{\ge 0}$ to a scalar value, $\phi \in \mathbb{R}$. The graph of this map, $(x,t,\phi(x,t))$, forms a 2-dimensional surface embedded in $\mathbb{R}^3$. For any fixed time $t_0$, the function $\phi(x,t_0)$ describes a curve in the $(x,\phi)$-plane.

The coalescence function $g$ is designed to activate (i.e., $g \approx 1$) in specific geometric configurations of this curve $\phi(x,t_0)$. It targets regions that represent a persistent, thin material layer (e.g., an interlayer in diffusion bonding, or the material between two approaching interfaces) which should not be eliminated by the phase-field dynamics unless specific physical conditions (e.g., via temperature-dependent thresholds) are met.

Conceptually, such a region (e.g., an interlayer represented by $\phi \approx 0$ when the bulk phases are $\phi \approx 1$) is characterized by $\phi(x,t)$ exhibiting a local spatial minimum at that value. The purpose of $g$ is to detect such minima on that surface and, if active, to counteract the thermodynamic driving force that would typically act to flatten this minimum and merge the adjacent bulk phases. It is crucial that $g$ robustly identifies these minima, distinguishing them from flat regions (where $\phi$ might be constant and small) or minor numerical oscillations. Small oscillations within a bulk phase (e.g., $\phi \approx 1$) should not trigger $g$ if the target value for interface arrest is, e.g., $\phi \approx 0$.

To formalize the detection of a local spatial minimum around a target value (e.g. $\phi \approx 0$), we impose the conditions summarized in Table~\ref{tab:1D_conditions} at any point $(x,t)$.
The first ensures that $\phi$ is near the target value characteristic of the layer to be preserved. 
The second identifies points in the vicinity of a local spatial extremum. 
The third ensures that this extremum is a minimum by requiring a sufficiently positive second derivative, distinguishing it from local maxima or flat regions.

\begin{table}[htb!]
    \centering
    \begin{tabular}{ll} 
    \toprule
    Criterion & Threshold Parameter(s) \\
    \midrule
     $0 \le \phi(x, t) < \alpha$        & $\alpha$ (small positive) \\
     $|\phi_x(x,t)| < \gamma$          & $\gamma$ (small positive) \\
     $\phi_{xx}(x,t) > \beta$          & $\beta$ (positive) \\
    \bottomrule
    \end{tabular}
    \caption{Geometric conditions for the coalescence kinetic function in 1D}
    \label{tab:1D_conditions}
\end{table}

The coalescence function is then defined as the product of regularized step functions representing these criteria:
\begin{equation}
    \label{eq:g_functional_1d}
    g(\phi, \phi_x, \phi_{xx}) = H_l(\alpha - \phi) \cdot H_l(\gamma - |\phi_x|) \cdot H_l(\phi_{xx} - \beta)
\end{equation}
where $H_l(u)$ is a regularized Heaviside step function; we use the form $H_l(u) = \frac{1}{2}(1+\tanh(u/l))$, where $l$ is a smoothing parameter.

This definition ensures that $g=1$ (interface arrest is fully active) if and only if all three conditions are satisfied simultaneously, thus precisely identifying the local spatial minima that represent the features whose evolution is to be controlled. If any condition is not met, $g=0$, and the standard kinetic evolution proceeds. The parameters $\alpha, \beta,$ and $\gamma$ are positive thresholds that determine the sensitivity and specificity of the interface detection. These parameters can be formulated to depend on various physical parameters such as temperature, stress state, or chemical potentials.
This structure provides a flexible framework for incorporating different physical mechanisms that might trigger interface merging, such as temperature-induced activation or stress-dependent behavior.

Importantly, even if $g$ becomes non-zero (e.g., $g=1$) at some points, if the local thermodynamic driving force, $\mathcal{F}$, happens to be zero at those points, the kinetic term in ~\ref{eq:our_evolution_hindered} will still be zero, and thus $\phi$ will not evolve artificially. The role of $g$ is to suppress evolution when there is a driving force that would otherwise act to coalesce interfaces that should not merge.

The definition of $g$ using nonlocal geometric properties of $\phi$ is a departure from using globally fixed numerical coefficients or relying solely on coupled fields (like concentration) to control interface stability. Instead, $g$ provides a dynamic, state-dependent response based on the instantaneous physical geometry of the phase distribution. The specific choice of these criteria and their extension to higher dimensions is the core of our proposed modification for modeling controlled interface evolution in processes like diffusion bonding.

\subsection{Geometric Control of Field Evolution in Higher Dimensions}

The principle of employing $g$, based on the nonlocal geometry of the phase field $\phi$, is extended now to higher-dimensional domains. This section first details the formulation for two-dimensional systems, leveraging principal curvatures, and subsequently outlines its generalization to three dimensions via Sylvester's criterion for assessing the positive-definiteness of the Hessian.

\subsubsection{Interface Detection via Principal Curvatures in Two Dimensions}

In a two-dimensional spatial domain $\Omega \subset \mathbb{R}^2$, the phase field $\phi(\bfx,t): \Omega \times [0,T] \to \mathbb{R}$ (for fixed $t$) defines a scalar field whose nonlocal geometric features can be used to identify specific interfacial regions. The function $g$ is formulated to activate in regions where $\phi$ exhibits a local spatial minimum, indicative of a persistent thin layer structure.

The characterization of such local minima involves an analysis of the Hessian matrix of second derivatives of $\phi$ with respect to the spatial coordinates $\bfx=(x_1,x_2)$:
\begin{equation}
    \mathcal{H}(\phi) := \nabla\nabla\phi = 
    \begin{bmatrix} 
        \phi_{x_1x_1} & \phi_{x_1x_2} \\ 
        \phi_{x_1x_2} & \phi_{x_2x_2} 
    \end{bmatrix}
\end{equation}
A point $\bfx$ corresponds to a strict local minimum of $\phi$ if, and only if, both principal curvatures, $k_1(\bfx)$ and $k_2(\bfx)$ (the eigenvalues of $\mathcal{H}(\phi)$), are positive. This condition ensures local convexity of the scalar field $\phi$. For two-dimensional systems, these principles are extended. The coalescence function $g$ is defined by the simultaneous satisfaction of the conditions detailed in Table~\ref{tab:2D_conditions}.

\begin{table}[htb!]
    \centering
    \begin{tabular}{ll}
    \toprule
    Criterion & Threshold Parameter(s) \\
    \midrule
     $0 \le \phi(\bfx,t) < \alpha$ & $\alpha$ (small positive) \\
     $|\nabla \phi(\bfx,t)| < \gamma$ & $\gamma$ (small positive) \\
     $k_1(\bfx,t) > \beta_{k1}$ & $\beta_{k1}$ (positive) \\
     $k_2(\bfx,t) > \beta_{k2}$ & $\beta_{k2}$ (positive) \\
    \bottomrule
    \end{tabular}
    \caption{Geometric conditions for activating the coalescence function in two dimensions.}
    \label{tab:2D_conditions}
\end{table}

The thresholds $\alpha, \gamma, \beta_{k1}, \beta_{k2}$ determine the sensitivity of the detection. The 2D coalescence function $g$ is then the product of these four regularized Heaviside functions, corresponding to the criteria in Table~\ref{tab:2D_conditions}:
\begin{equation}
\label{eq:g_functional_2d_principal_curvatures_final}
    g(\phi, \nabla \phi, k_1, k_2) = H_l(\alpha - \phi) \cdot H_l(\gamma - |\nabla \phi|) \cdot H_l(k_1 - \beta_{k1}) \cdot H_l(k_2 - \beta_{k2})
\end{equation}

Activation of $g$ (that is, $g \approx 1$) occurs only when all criteria are met, thus identifying local minima of $\phi$ targeted for controlled evolution. The term ``interface detection'' herein refers to the identification of these specific features.

Although the formulation above is expressed in terms of the principal curvatures $(k_1,k_2)$, in finite element computations we enforce the same geometric conditions through the invariants of the Hessian:
\begin{equation}
    \label{eq:g_functional_2d_principal_curvatures_invariants}
    \operatorname{tr}(\mathcal{H}(\phi))=k_1+k_2,\qquad
    \det(\mathcal{H}(\phi))=k_1k_2
\end{equation}
These quantities are more stable to evaluate numerically.

Accordingly, the conditions $k_1 > 0$ and $k_2 > 0$ are realized by the filtered inequalities
\begin{equation}
 H_l(\operatorname{tr}(\mathcal{H}(\phi))-\beta_{L}) \quad\text{and}\quad 
 H_l(\det(\mathcal{H}(\phi))+\beta_{D})
\end{equation}
with small tolerances $\beta_{L},\beta_{D}>0$ chosen to admit nearly flat minima ($k_2\approx 0$) while excluding saddle regions ($k_1k_2<0$). This ensures that the theoretical curvature criteria of Table~\ref{tab:2D_conditions} are faithfully captured in the discrete implementation.

\begin{figure}[htb!]
    \centering
    \includegraphics[width=0.6\textwidth]{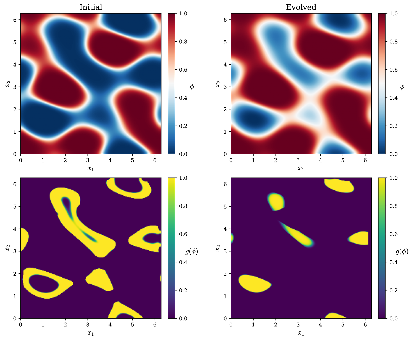} 
    \caption{Comparison of a scalar field \(\phi(x_1,x_2)\) (top row) and the corresponding activation of the function \(g(\phi)\) (bottom row), before (left) and during (right) a potential coalescence event. The function \(g\) localizes to regions satisfying the geometric criteria of a persistent minimum.}
    \label{fig:phi-interface-comparison}
\end{figure}

Figure~\ref{fig:phi-interface-comparison} illustrates this mechanism. The top row shows the scalar field $\phi(x_1,x_2)$ at two stages of an evolution where coalescence might be expected. The bottom row shows the corresponding activation regions $g(\phi)$. As $\phi$ evolves, the regions where $g=1$ persist where the geometric criteria for a local minimum are met, thereby suppressing the kinetic term in the evolution equation (e.g. \ref{eq:our_evolution_hindered}) and preventing the elimination of the interlayer.

\subsubsection{Generalization to Three Dimensions using Sylvester's Criterion}

For a three-dimensional spatial domain, while the condition of all positive principal curvatures remains the theoretical basis for identifying a local minimum, Sylvester’s criterion offers a direct algebraic test for the positive-definiteness of the $3 \times 3$ Hessian matrix $\mathcal{H}(\phi):=\nabla\nabla\phi$. $\mathcal{H}(\phi)$ is positive-definite if and only if all of its leading principal minors, $\Delta_k$, are positive \cite{carmo_differential_2016, horn_matrix_2012}. This provides a robust framework for higher dimensions.

The 3D coalescence function, $g_{\text{3D}}$, thus incorporates these leading principal minor conditions:
\begin{equation}
    g_{\text{3D}}(\phi, \nabla\phi, \nabla\nabla\phi) = 
    H_l(\alpha-\phi) \cdot 
    H_l(\gamma-|\nabla\phi|) \cdot 
    \prod_{k=1}^{3} H_l\bigl(\Delta_{k}(\mathcal{H}(\phi))-\beta_{k}\bigr)
    \label{eq:g3D_final}
\end{equation}
where $\Delta_{k}(\mathcal{H}(\phi))$\footnote{
    The leading principal minors are defined as: $\Delta_{1}=\phi_{x_1x_1}$, $\Delta_{2}= \begin{vmatrix}\phi_{x_1x_1}&\phi_{x_1x_2}\\[0.5ex]\phi_{x_1x_2}&\phi_{x_2x_2}\end{vmatrix}$, and $\Delta_{3}=\det(\mathcal{H}(\phi))$. While the values of these specific minors are coordinate-dependent, the condition that they are all simultaneously positive is equivalent to the coordinate-invariant statement that all eigenvalues of the Hessian (the principal curvatures) are positive. For $g_{\text{3D}}$ to activate, the conditions $\Delta_1 > \beta_1$, $\Delta_2 > \beta_2$, and $\Delta_3 > \beta_3$ must hold.
} is the $k$-th leading principal minor, and $\beta_{k}$ are small positive thresholds. This formulation ensures $g_{\text{3D}} \approx 1$ exclusively at true local spatial minima of $\phi$, reflecting the local convexity of the scalar field.

\begin{figure}[htb!]
    \centering
        \subfloat[3D visualization of a spherical inclusion ($\phi=1$ inside, $\phi=0$ outside) with a cutaway revealing the interior.]{%
        \includegraphics[width=0.5\textwidth]{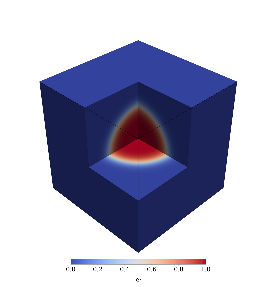}
        \label{fig:3d_cutaway}%
    }
    \vspace{1em} 
    \subfloat[Cross-sectional views showing the phase field $\phi$ (top row of each slice) and the activation of the interface detection function $g_{3D}(\phi)$ (bottom row of each slice) at three different depths.]{
        \includegraphics[width=0.75\textwidth]{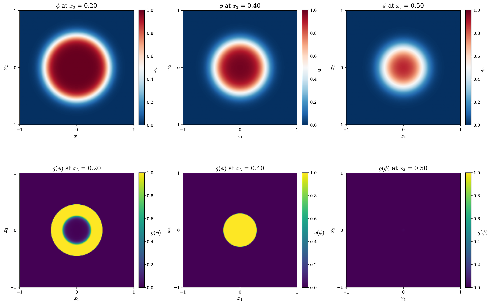}
        \label{fig:3d_slices}%
    }
    \caption{Visualization of a 3D phase field with interface detection based on Sylvester's criterion. The function $g_{3D}$ accurately identifies the interior region where $\phi$ approaches a local minimum.}
    \label{fig:3d_visualization}
\end{figure}

Figure~\ref{fig:3d_visualization} illustrates this 3D interface detection for a spherical inclusion test case. Sylvester's criterion ensures that the coalescence functional accurately identifies regions of strict local minima, providing robust control over interface evolution in full three-dimensional simulations.

\subsection{Physical Interpretation and Connection to Sharp Interface Limits}

The geometric criteria employed in $g$ have connections to quantities that emerge naturally in the sharp interface limit of phase-field models \cite{storvik_sharp-interface_2025}. 
In particular, the Laplacian and determinant of the Hessian relate directly to the curvature terms that govern interface dynamics in the sharp interface limit. 
As demonstrated by Storvik and Bringedal \cite{storvik_sharp-interface_2025}, when applying formal matched asymptotic expansions to phase-field equations, the chemical potential at the interface is influenced by curvature and elastic energy contributions.
The parameters $\alpha$, $\beta$, and $\gamma$ can be interpreted as physically meaningful thresholds rather than merely numerical constants: they determine when the geometry creates conditions for interface preservation versus coalescence, reflecting the underlying thermodynamic state of the system. 
These thresholds effectively capture the critical conditions under which distinct interfaces should remain separate, a key requirement in modeling diffusion bonding processes like those studied by Kovacevic et al.~\cite{kovacevic_interfacial_2020}.

This approach differs from previous phase-field formulations for diffusion bonding. 
In contrast to models that rely primarily on solute concentration fields to drive interface evolution, our method implements geometric control over interface dynamics through curvature-related invariants. This enables us to capture scenarios where interfaces should remain distinct despite close proximity, a common requirement in diffusion bonding processes, where coalescence may be thermodynamically or kinetically unfavorable under certain conditions.
The use of such geometric measures aligns with concepts explored in phase-field theory, where differential geometric quantities can arise in asymptotic analyses and have been linked to physical interface properties or effective interface Hamiltonians \cite{sekerka_morphology_2004, elder_sharp_2001,sun_sharp_2007}.

\subsection{Compatibility with the Second Law of Thermodynamics}

The thermodynamic admissibility of the proposed phase-field model is established by demonstrating its consistency with the second law of thermodynamics. In an isothermal setting, this requires that the rate of energy dissipation be non-negative for any process occurring within the system. Following the standard process \cite{penrose_thermodynamically_1990,karimi_energetic_2022, de_macedo_disclinations_2018, agrawal_dependence_2017,naghibzadeh_accretion_2025,chua_phase-field_2022,KARIMI2025106232,chua2024deformation}, we compute the time derivative of $\mathcal{P}$ (\ref{eq:free_energy}), to get:
\begin{equation}
    \deriv{\mathcal{P}}{t} = \int_\Omega \variation_\phi\mathcal{P} \frac{\partial \phi}{\partial t} \dm V
\end{equation}
Substituting \eqref{eq:our_evolution_hindered} into the above yields:
\begin{equation}
\label{eq:dissipation_final_example}
    \deriv{\mathcal{P}}{t}     
    = \int_\Omega \variation_\phi \mathcal{P} \left[ -\kappa (1-g)\variation_\phi \mathcal{P} |\nabla \phi| \right] \dm V
    = -\int_\Omega \kappa (1-g) \left(\variation_\phi \mathcal{P}\right)^2 |\nabla \phi| \dm V
\end{equation}
For the second law to be satisfied $\left( \deriv{\mathcal{P}}{t}  \leq 0\right)$, the integrand in \eqref{eq:dissipation_final_example} must be non-negative. Given that $\kappa > 0$, $|\nabla \phi| \ge 0$, and $\left(\variation_\phi \mathcal{P}\right)^2 \ge 0$, the condition for non-positive $\deriv{\mathcal{P}}{t} $ reduces to requiring $1-g \geq 0$. As $g$ is defined to be within the interval $[0, 1]$, thermodynamic admissibility is satisfied.
Thus, the incorporation of the coalescence function $g$ allows for localized modulation of the phase-field kinetics---effectively slowing or arresting evolution in specific regions---while ensuring that the total free energy does not increase, thereby maintaining consistency with the second law of thermodynamics. 

\subsection{Mixed Finite Element Implementation}
The governing phase-field evolution equation \eqref{eq:our_evolution_hindered} incorporates $g$, which is a function of nonlocal geometric measures including second-order spatial derivatives of the phase field $\phi$. The presence of such higher derivatives, potentially in nonlinear combinations with first-order terms (e.g., $|\nabla\phi|$), alongside the Laplacian $\nabla^2\phi$ in the thermodynamic driving force, requires a mixed finite element formulation, as the geometric hindering term $g(\phi,\nabla\phi,\nabla^2\phi,\dots)$ introduces both first and second derivatives whose consistent evaluation in weak form necessitates explicit access to $\nabla\phi$ and $\nabla\nabla\phi$.

We introduce an auxiliary vector field $\bfpsi := \nabla\phi$ as an independent unknown \cite{brezzi_mixed_1991}. This approach recasts the problem into a system of coupled, first-order partial differential equations by providing direct access to both $\nabla\phi$ and $\nabla\nabla\phi$, which is essential for evaluating the geometric hindering contributions. While we have implemented simpler discretization strategies like finite differences in one-dimensional contexts, the mixed FEM offers enhanced flexibility and robustness for multi-dimensional problems with unstructured meshes. 
This mixed formulation is implemented using FEniCS \cite{logg_automated_2012, alnaes_fenics_2015}, a versatile open-source platform for solving partial differential equations.

To establish the mixed variational formulation over a domain $\Omega \subset \mathbb{R}^d$, the phase field $\phi$ is sought in the Sobolev space $V^\phi = H^1(\Omega)$, and the auxiliary gradient field $\bfpsi$ is sought in $V^{\bfpsi} = H(\mathrm{div}; \Omega)$. Specifically, $V_h^\phi$ is constructed using second-order (P2) Continuous Galerkin (CG) finite elements, which employ $C^0$-continuous quadratic Lagrange basis functions. For $V_h^{\bfpsi}$, we employ first-order (P1) Continuous Galerkin vector elements; this choice is formally conforming as the space of such $C^0$-continuous vector elements is a subspace of $(H^1(\Omega))^d$, and subsequently of $H(\mathrm{div}; \Omega)$, ensuring a well-defined divergence. 

The mixed weak formulation is then: find the discrete fields $(\phi_h, \bfpsi_h) \in V_h^\phi \times V_h^{\bfpsi}$ such that for all corresponding test functions $(\hat{\phi}_h, \hat{\bfpsi}_h) \in V_h^\phi \times V_h^{\bfpsi}$, the following equations hold. If the system is coupled with mechanics, a discrete displacement field $\bfu_h \in V_h^{\bfu}$ (where $V_h^{\bfu}$ is typically a subspace of $(H^1(\Omega))^d$) and its corresponding test function $\hat{\bfu}_h \in V_h^{\bfu}$ are also included (See Section \ref{sec:mechanical}).
\begin{equation}
    \int_{\Omega} \bfpsi_h \cdot \hat{\bfpsi}_h \, \dm V - \int_{\Omega} \nabla \phi_h \cdot \hat{\bfpsi}_h \, \dm V = 0
    \label{eq:weak_form_psi_definition_final}
\end{equation}
and
\begin{equation}
    \int_{\Omega} \frac{\phi_h^{n+1} - \phi_h^n}{\Delta t} \hat{\phi}_h \, \dm V - \int_{\Omega} \mathcal{R}_{\phi}(\phi_h^{n+\theta}, \bfpsi_h^{n+\theta}, g_h^n) \hat{\phi}_h \, \dm V = 0
    \label{eq:weak_form_phi_evolution_main_final}
\end{equation}

The weak form comprises \eqref{eq:weak_form_psi_definition_final}, which enforces the definition of the auxiliary gradient field $\bfpsi_h = \nabla\phi_h$ in a weak sense, and \eqref{eq:weak_form_phi_evolution_main_final}. 
In the latter, the term $\mathcal{R}_{\phi}$ represents the spatial operator obtained from \eqref{eq:our_evolution_hindered}, which incorporates the geometric term. 
In particular, the contributions involving $|\nabla\phi|$ and $\nabla^{2}\phi$ generate weak-form terms of the type $\varepsilon\,\nabla\phi_h^{n+1}\!\cdot\nabla(\hat\phi_h\,m_n)$, which arise from integration by parts of the Laplacian $\varepsilon m \nabla^{2}\phi$; here, $m = |\nabla\phi| (1-g)$ is a function of $\phi$ but is treated as a known value from the previous step.
The mixed formulation provides consistent approximations to both $\nabla\phi$ and $\nabla\nabla\phi$ which are required to evaluate the geometric function $g$.

For the temporal discretization, evaluating the term $g_h^n = g(\phi_h^n, \bfpsi_h^n, \nabla  \bfpsi_h^n, \dots)$ using values from the previous time step ($n$) renders the scheme semi-implicit when terms involving $\phi_h^{n+\theta}$ and $\bfpsi_h^{n+\theta}$ (with $\theta > 0$, e.g., $\theta=1$ for Backward Euler or $\theta=0.5$ for Crank-Nicolson) are treated implicitly. This choice is favored over fully explicit schemes due to the inherent stiffness associated with phase-field equations, which can impose severe restrictions on the time step size for stability \cite{boscarino_implicit-explicit_2013, hundsdorfer_numerical_2013, ascher_implicit-explicit_1995}.

In addition to being evaluated explicitly from the previous iterate, the hindering field is updated in a strictly monotone manner. Let $g_{\mathrm{inst}}^{n}$ denote the instantaneous geometric hindering computed from $(\phi_h^{n}, \bfpsi_h^{n}, \nabla\bfpsi_h^{n})$. The stored hindering field $g_{\mathrm{pers}}^{n}$ is then advanced by
\begin{equation}
g_{\mathrm{pers}}^{\,n+1}
   = \max\!\left(g_{\mathrm{pers}}^{\,n},\, g_{\mathrm{inst}}^{\,n}\right),
\label{eq:g_persistence_update_short}
\end{equation}
so that the effective hindering factor used in the mobility $m_n = |\nabla\phi_h^n|\,(1-g_{\mathrm{pers}}^{\,n})$ cannot decrease once activated. This construction parallels the irreversibility condition in phase-field fracture that prevents the unphysical healing of cracks \cite{hakimzadeh_phase-field_2025}.

The discretization of \eqref{eq:weak_form_psi_definition_final} and \eqref{eq:weak_form_phi_evolution_main_final} at each time step $\Delta t$ yields a coupled system of non-linear algebraic equations for the discrete unknowns $(\phi_h^{n+1}, \bfpsi_h^{n+1})$. This system is solved using a Newton-Raphson iterative method.

\section{Numerical Results}
\label{sec:numericalresults}

This section presents numerical simulations to demonstrate the capabilities of the proposed phase-field model, with a particular focus on the role of $g$ in controlling interface coalescence. First, a one-dimensional example will illustrate the interface detection criteria and their effect on the evolution of the phase field. Subsequently, the behavior will be explored in several two-dimensional problems.

\subsection{1D Simulations: Demonstration of Coalescence Suppression}
\label{ssec:1d_case}

The 1D simulations contrast the conventional kinetic behavior leading to interface merging with the controlled evolution achieved by incorporating the coalescence term.

The evolution of the phase field $\phi(x,t)$ in the 1D simulations is governed by \eqref{eq:1d_hindered_evolution_specific}, utilizing the 1D function $g=g(\phi, \phi_x, \phi_{xx})$ as defined in \eqref{eq:g_functional_1d} and incorporating the kinetic form from \cite{agrawal_dynamic_2015-1}. This governing equation was discretized using an explicit finite difference scheme for both spatial and temporal derivatives. For the simulations presented, the spatial step was $\Delta x = 0.01$, the time step was $\Delta t = 1 \times 10^{-4}$, the gradient energy coefficient $\kappa_\phi = 0.1$, and the smoothing parameter for the Heaviside functions within $g$ was $l = 0.02$. The coalescence thresholds were set to $\alpha=0.2$, $\beta=0.1$, and $\gamma=0.05$. The domain was of length $L=2.0$ with Dirichlet boundary conditions ($\phi=1$) at $x=0$ and $x=L$.

The initial condition consisted of two regions where $\phi=1$, separated by a region where $\phi=0$. Specifically, $\phi(x,0)=1$ for $x \le 0.5$ and $x \ge 1.5$, and $\phi(x,0)=0$ for $0.5 < x < 1.5$, with smooth transitions at the interfaces.

\begin{figure}[htb!]
    \centering
    \subfloat[Conventional evolution ($g=0$).]{
        \includegraphics[width=0.48\textwidth]{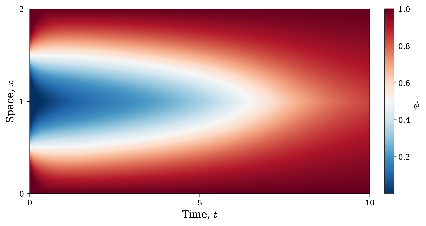} 
        \label{fig:1d_conventional}
    }
    \hfill
    \subfloat[Evolution with active coalescence suppression ($g \neq 0$).]{
        \includegraphics[width=0.48\textwidth]{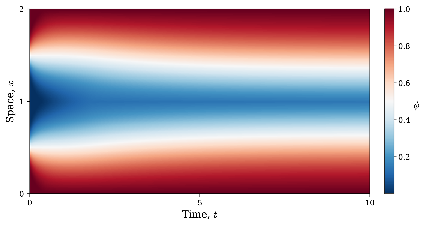} 
        \label{fig:1d_hindered}
    }
    \caption{Space-time contour plots of $\phi(x,t)$ evolution in 1D. (a) Conventional model where interfaces merge, and the domain becomes $\phi \approx 1$. (b) With the coalescence functional active, interfaces approach but a stable $\phi \approx 0$ region is preserved.}
    \label{fig:1d_evolution_comparison}
\end{figure}

Figure~\ref{fig:1d_evolution_comparison}(a) illustrates the conventional evolution scenario, achieved by setting $g=0$ in \eqref{eq:agrawal_evolution}. As expected, the two interfaces (boundaries between $\phi \approx 1$ and $\phi \approx 0$) move towards each other due to the driving force seeking to minimize the total interface length and gradient energy. Over time, these interfaces coalesce, and the central $\phi \approx 0$ region is eliminated, resulting in the entire domain uniformly approaching $\phi=1$.

In contrast, Figure~\ref{fig:1d_evolution_comparison}(b) shows the evolution when $g$ is active. Initially, the interfaces move inwards, similar to the conventional case. However, as the central $\phi \approx 0$ region narrows and the geometric conditions for $g$ (i.e., $\phi < \alpha$, $|\phi_x| < \gamma$, and $\phi_{xx} > \beta$) are met within this region, $g$ approaches $1$. This significantly reduces the effective mobility $(1-g)$, thereby halting further inward motion of the interfaces. Consequently, a stable region of $\phi \approx 0$ is preserved between the two $\phi \approx 1$ walls, demonstrating the suppression of coalescence.

A crucial aspect of this coalescence mechanism is its targeted action. The activation of $g$ (i.e., $g \approx 1$) effectively nullifies the kinetic prefactor in \eqref{eq:1d_hindered_evolution_specific}, thereby counteracting the thermodynamic driving force $\left( \kappa_\phi\phi_{xx} - 2 \phi(2 \phi-1)(\phi-1) \right)$ that would otherwise cause merging. Importantly, if the geometric conditions for $g$ were to be met in a region where this thermodynamic driving force is already zero (e.g., in a flat, equilibrium part of the $\phi$ field far from the interfaces), the activation of $g$ would not artificially induce any evolution, as $\frac{\partial\phi}{\partial t}$ would remain zero. This ensures that the coalescence function only intervenes when necessary to prevent unphysical interface elimination.

\subsection{2D Simulations: Controlled Interface Evolution}
\label{ssec:2dcase}

Three sets of two-dimensional simulations were conducted to illustrate the efficacy in managing interface coalescence dynamics under different geometric configurations: (i) initially parallel vertical interfaces, (ii) initially parallel diagonal interfaces, and (iii) a circular inclusion. Unless otherwise specified, all simulations utilized a rectangular domain of size $2.0 \times 1.0$, discretized with uniform finite elements of characteristic size $\Delta x = \Delta y = 0.02$; we have checked through mesh refinement for selected simulations that the results are independent of the discretization.
The gradient energy coefficient was set to $\kappa_\phi=0.1$, yielding a ratio of $\kappa_\phi / \Delta x=5$. Simulations were run until the interface motion effectively ceased (about $t = 1.0$). The parameters for the function $g$ (i.e., $\alpha, \beta, \gamma$) were chosen appropriately for each case to demonstrate controlled hindrance, with specific values detailed in Section \ref{sec:sensivity_analysis}.

\subsubsection{Parallel Vertical Interfaces}
\label{sec:vertical-walls}

\begin{figure}[htb!]
    \centering
    \includegraphics[width=0.48\textwidth]{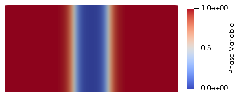}
    \caption{Initial condition for the parallel vertical interfaces case.}
    \label{fig:verticalWallsInitial}
\end{figure}

The initial condition comprises two regions where $\phi = 1$ ($x \le 0.8$ and $x \ge 1.2$), separated by a region where $\phi = 0$, as shown in Figure~\ref{fig:verticalWallsInitial}. 
Homogeneous Neumann boundary conditions ($\nabla\phi \cdot \bfn = 0$) were applied at the top and bottom boundaries, while Dirichlet conditions ($\phi = 1$) were imposed on the left and right boundaries. 

\begin{figure}[htb!]
    \centering
    \subfloat[Equilibrium $\phi$]{
        \includegraphics[width=0.45\textwidth]{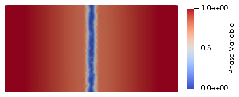}
        \label{fig:verticalWallsPhi}
    }
    \hfill
    \subfloat[Extracted interface ($g$ contour)]{
        \includegraphics[width=0.45\textwidth]{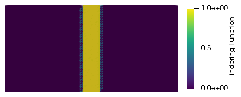}
        \label{fig:verticalWallsInterface}
    }
    \caption{Equilibrium state for the parallel vertical interfaces, showing a persistent gap. (a) Phase-field $\phi$. (b) Interface contour.}
    \label{fig:verticalWallsResults}
\end{figure}

Figure~\ref{fig:verticalWallsResults} illustrates a typical final state where $g$ has arrested the inward motion of the interfaces, preserving a distinct gap. The width of this gap is influenced by the choice of coalescence parameters and numerical resolution, as discussed below.

\subsubsection{Parallel Diagonal Interfaces}
\label{sec:diagonal-walls}

\begin{figure}[htb!]
    \centering
    \includegraphics[width=0.48\textwidth]{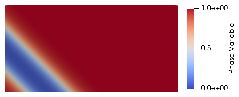}
    \caption{Initial condition for the parallel diagonal interfaces case.}
    \label{fig:diagWallsInitial}
\end{figure}

In this configuration, the region where $\phi = 1$ forms a diagonal band across the domain (Figure~\ref{fig:diagWallsInitial}). Dirichlet conditions ($\phi = 1$) were set along the top and right edges, with no-flux conditions on the other boundaries. The initial interfaces were again smoothly defined.

\begin{figure}[htb!]
    \centering
    \subfloat[Equilibrium $\phi$]{
        \includegraphics[width=0.45\textwidth]{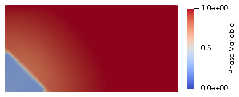}
        \label{fig:diagWallsPhi}
    }
    \hfill
    \subfloat[Extracted interface ($g$ contour)]{
        \includegraphics[width=0.45\textwidth]{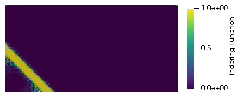}
        \label{fig:diagWallsInterface}
    }
    \caption{Equilibrium state for the parallel diagonal interfaces, showing a persistent diagonal gap. (a) Phase-field $\phi$. (b) Interface contour.}
    \label{fig:diagWallsResults}
\end{figure}

The evolution demonstrates the ability to handle non-axis-aligned interfaces. Figure~\ref{fig:diagWallsResults} shows a representative final state where the coalescence term maintains a distinct diagonal interface, preventing the complete coalescence of interior interfaces, even when the lower branch (adjacent to the Neumann boundaries) delays merging and eventually disappears.

\subsubsection{Circular Inclusion Structure}
\label{sec:elliptical-structure}

\begin{figure}[ht!]
    \centering
    \includegraphics[width=0.48\textwidth]{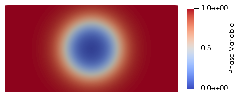}
    \caption{Initial condition for the circular inclusion structure.}
    \label{fig:ellipseInitial}
\end{figure}

This case has an initial inclusion region with $\phi = 0$ within a matrix with $\phi = 1$ (Figure~\ref{fig:ellipseInitial}). 
Dirichlet conditions ($\phi = 1$) were applied along the left and right boundaries to partially pin the structure, while no-flux conditions were used elsewhere, allowing the unpinned portions of the ellipse to evolve. For this configuration, a finer mesh ($\Delta x = 0.01$) has been used to provide  better convergence.

\begin{figure}[h!]
    \centering
    \subfloat[Equilibrium $\phi$]{
        \includegraphics[width=0.45\textwidth]{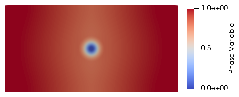}
        \label{fig:ellipsePhi}
    }
    \hfill
    \subfloat[Extracted interface ($g$ contour)]{
        \includegraphics[width=0.45\textwidth]{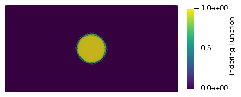}
        \label{fig:ellipseInterface}
    }
    \caption{Equilibrium state for the elliptical interface structure. (a) Phase-field $\phi$. (b) Interface contour.}
    \label{fig:ellipseFinal}
\end{figure}

Curvature-driven evolution, which typically leads to the shrinkage and eventual disappearance of such an inclusion in standard phase-field models, is modulated here by $g$. Figure~\ref{fig:ellipseFinal} shows a final state in which the elliptical structure is preserved, although its exact shape and size depend on the coalescence parameters and other numerical parameters. This case highlights the ability to stabilize non-trivial curved interfaces.

\subsection{Parametric Sensitivity}
\label{sec:sensivity_analysis}

The numerical solutions, particularly the efficacy of the coalescence kinetic function $g$, are influenced by the numerical discretization parameters ($\Delta x$, $\Delta t$) and parameters within the definition of $g$, such as the regularization parameter $l$ in the step-like functions and the geometric thresholds ($\alpha$, $\beta$ ,$\gamma$). 
A systematic study was performed to quantify these sensitivities. The parameters for $g$ in the Heaviside functions were: $\alpha = 0.2$, $\gamma = 1.5$, and $\beta_{k1} = \beta_{k2} = 0.1$ (for 2D principal curvatures), unless specified otherwise during the sensitivity study of these particular thresholds. The regularization parameter for the step functions was set to  $2 \Delta x\leq l$.

\subsubsection{Influence of Numerical Discretization Parameters}

\begin{figure}[htb!]
    \centering
    \subfloat[Coarser mesh: $\Delta x \approx 0.02$]{
        \includegraphics[width=0.47\textwidth]{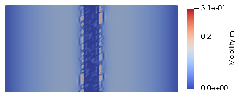}
        \label{fig:coarse_mesh}
    }
    \hfill
    \subfloat[Finer mesh: $\Delta x \approx 0.01$]{
        \includegraphics[width=0.47\textwidth]{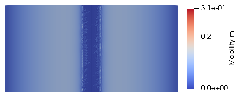}
        \label{fig:fine_mesh}
    }
    \caption{Effect of mesh resolution on the evolution for the vertical interface case at $t=4.0$ with $\Delta t = 2 \times 10^{-4}$, $l=0.05 \approx 2.5 \Delta x_{\text{coarse}}$.}
    \label{fig:mesh_comparison}
\end{figure}

The choice of spatial mesh size ($\Delta x$) and time step size ($\Delta t$) impacts solution accuracy and stability, as is standard for numerical solutions of PDEs.
\begin{itemize}

    \item \textbf{Spatial Resolution ($\Delta x$):} 
    Simulations were performed with varying mesh resolutions. In addition to examining the phase field $\phi$, we also visualize the effective mobility field $m = |\nabla\phi| (1-g)$, which highlights where the interface is still actively evolving. 
    As shown in Figure~\ref{fig:mesh_comparison}, the finer mesh produces a sharper and more localized band of high mobility near the interlayer interfaces, whereas the coarser mesh yields a broader and more diffuse active region.
    To allow direct comparison, both panels use the same color scale, normalized to the maximum $m$ value obtained on the finer mesh.
    While the general behavior of interface arrest is preserved across mesh sizes, the finer mesh equilibrates to a slightly lower mean $\phi$ in the center of the interlayer, reflecting improved resolution of the balance between chemical, gradient, and elastic energies.
    As before, the precise width and profile of the arrested interface can exhibit mesh dependence if the diffuse interface (proportional to $\kappa_\phi$) is under-resolved (i.e., if $\kappa_\phi/\Delta x$ is too small, less than~$2$). For all results presented, care was taken to ensure sufficient resolution.
    
    \item \textbf{Time Step Size ($\Delta t$):} 
    To assess temporal discretization effects, we compare two simulations of the diagonal interface case: one using the default time step $\Delta t = 2\times10^{-4}$, and another using a larger step $\Delta t = 2\times10^{-3}$.
    The smaller time step resolves sharper local variations in the mobility field---in particular the fine oscillations of $m = |\nabla\phi|(1-g)$ near the evolving interface---while the larger step advances more diffusively, smoothing over these rapid transients.
    Despite these differences in intermediate evolution, both simulations converge to essentially the same arrested interface with $\phi \approx 0.26$.
    This demonstrates that the semi-implicit scheme is temporally stable over an order-of-magnitude change in $\Delta t$, and that moderately larger time steps can be used to improve computational efficiency without affecting the final equilibrium state.
\end{itemize}

\begin{figure}[htb!]
    \centering
    \subfloat[Smaller time step: $\Delta t = 2\times10^{-4}$]{
        \includegraphics[width=0.47\textwidth]{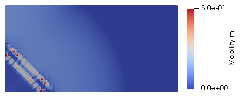}
        \label{fig:small_dt}
    }
    \hfill
    \subfloat[Larger time step: $\Delta t = 2\times10^{-3}$]{
        \includegraphics[width=0.47\textwidth]{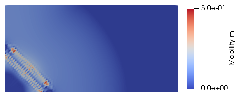}
        \label{fig:large_dt}
    }
    \caption{Effect of time step size ($\Delta t$) on the evolution for the diagonal interface case, at $t=4.0$ with $\Delta x \approx 0.02$, $l=0.05$.}
    \label{fig:diag_overshooting_comparison}
\end{figure}

\subsubsection{Influence of Coalescence Function Parameters}

\begin{figure}[htb!]
    \centering
    \subfloat[$l \approx 1.0\,\Delta x$]{
        \includegraphics[width=0.47\textwidth]{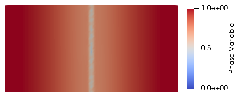}
        \label{fig:small_delta}
    }
    \hfill
    \subfloat[$l \approx 2.5\,\Delta x$]{
        \includegraphics[width=0.47\textwidth]{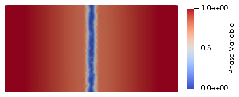}
        \label{fig:large_delta}
    }
    \caption{Effect of the Heaviside regularization parameter $l$ on the equilibrium state.}
    \label{fig:delta_comparison}
\end{figure}

The parameters within the function $g$ directly control its activation and thus the interface dynamics.
\begin{itemize}
    \item \textbf{Regularization Parameter in Step Functions:} 
    The parameter $l$ controls the sharpness and spatial extent of the regularized Heaviside functions used in $g$. A smaller $l$ (e.g., $l \approx \Delta x$) produces a very sharp transition, so the kinetics due to $g$ activates only in a narrow band: the diffuse interface is well resolved and becomes very steep, and the minimum value of $\phi$ at the mid–gap rises above $\phi=0.5$ relatively quickly as the band fills in. In contrast, a larger $l$ (e.g., $l \approx 2.5\,\Delta x$) smooths the transition so that $g$ acts over a thicker region; this spreads the hindering effect, slows the filling of the gap, and leads to a wider, more persistent low–$\phi$ band, as seen in Figure~\ref{fig:delta_comparison}.

    \item \textbf{Geometric Thresholds ($\alpha, \beta, \gamma$):} These thresholds determine the sensitivity of the coalescence function to the nonlocal geometry of the phase field. Varying these parameters allows for tuning the conditions under which interface coalescence is arrested. For instance, a smaller $\alpha$ or larger $\beta$ values make the coalescence criteria stricter, leading to earlier arrest and wider gaps. These parameters are intended to be physically motivated, representing, for example, critical curvatures or proximity measures below which merging is thermodynamically or kinetically unfavorable. A brief summary of representative parameter variations and their impact on interface evolution is presented in Table~\ref{tab:hindering_param_sensitivity_summary}.
\end{itemize}

\begin{table}[htb!]
    \centering
    \begin{tabular}{lccc}
    \hline
    \textbf{Parameter} & \thead{\textbf{Change in} \\ \textbf{Parameter Value}} & \thead{\textbf{Onset of Arrest} \\ \textbf{($g \approx 1$)}} & \thead{\textbf{Resultant Preserved} \\ \textbf{Interlayer Width}} \\
    \hline
    $\alpha$    & $\downarrow$  & \makecell{Later}  & Narrower \\
                                  & $\uparrow $    & \makecell{Earlier }    & Wider \\
    \hline
    $\beta$ & $\uparrow$   & \makecell{Later}  & Narrower \\
                                  & $\downarrow $   & \makecell{Earlier }    & Wider \\
    \hline
    $\gamma$ & $\downarrow $  & \makecell{Later}  & Narrower \\
                                  & $\uparrow $    & \makecell{Earlier }    & Wider \\
    \hline
    \end{tabular}
    \caption{Qualitative sensitivity to coalescence thresholds; $\uparrow$ denotes an increase in the parameter value, $\downarrow$ denotes a decrease.}
    \label{tab:hindering_param_sensitivity_summary}
\end{table}

Overall, these studies indicate that while numerical parameters (\(\Delta x, \Delta t\)) require careful selection for convergence and stability (as is standard), the parameters of the function $g$ (\(l, \alpha, \beta, \gamma\)) provide direct physical parameters to control the desired interface behavior, allowing the model to be calibrated to specific material systems or bonding conditions.


\section{Diffusion Bonding in Ti-ZrC Coupled to Mechanics}
\label{sec:mechanical}

Experimental studies on the diffusion bonding of ZrC-SiC composite with a Titanium (Ti) interlayer have demonstrated that the efficacy of the joining process \cite{lin_rapidly_2023}, particularly the homogenization of the joint, is critically dependent on the initial thickness of the Ti interlayer. Kovacevic et al \cite{kovacevic_interfacial_2020} addressed this system using a phase-field model coupled with a Cahn-Hilliard equation for carbon concentration. Their work identified the dependence of interfacial energy on the carbon concentration jump across the interface as a key thermodynamic driver, successfully predicting a critical interlayer thickness. 

\begin{figure}[htb!]
     \centering
     \includegraphics[width=0.4\textwidth]{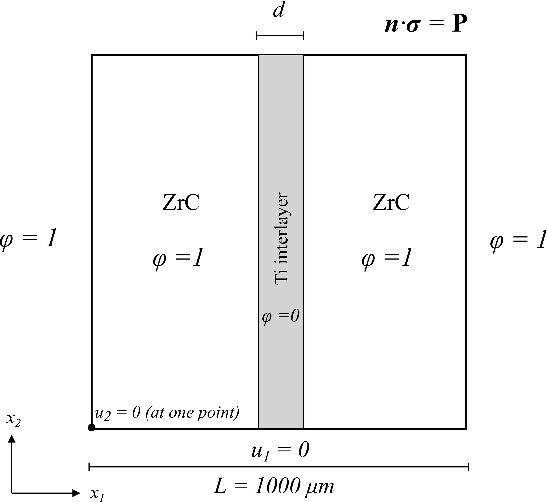}
     \caption{Schematic of the Ti-ZrC system. The region where $\phi = 0$ represents the Ti interlayer, and $\phi = 1$ represents the ZrC. The mechanical boundary conditions consist of a uniform traction corresponding to a compressive pressure of $20 MPa$ applied to all external boundaries. For the phase-field, we set $\phi=1$ at the right and left boundaries to represent the bulk ZrC blocks, with Neumann conditions elsewhere.}
     \label{fig:Ti-ZrC_setup_example}
\end{figure}

In contrast, this work proposes an alternative phase-field approach to model the same system without explicitly resolving concentration fields. 
The interplay of thermodynamics and kinetics leading to phenomena such as the emergence of a critical interlayer thickness can be effectively modeled by making the term $g$ -- through the threshold parameters $\alpha, \beta, \gamma$ -- dependent on macroscopic variables such as temperature. 
This allows the model to tune interface coalescence response based on prescribed thermodynamic conditions (e.g., a temperature threshold for bonding that might be linked to a critical carbon activity) without the computational expense of solving an additional advection-diffusion equation for concentration.
This approach focuses on capturing the macroscopic outcome of the bonding process as controlled by the phase-field evolution. Figure~\ref{fig:Ti-ZrC_setup_example} illustrates the representative computational domain for simulating the Ti interlayer ($\phi \approx 0$) between ZrC blocks ($\phi \approx 1$).

The model incorporates mechanical effects through an elastic energy density $U(\nabla \bfu, \phi)$, where $\bfu$ is the displacement, and phase evolution through a chemical free energy $W(\phi)$ and a phase-field gradient energy term. The total free energy functional $\mathcal{P}$ for this system, omitting explicit concentration dependence, is:
\begin{equation}
\mathcal{P} = \int_\Omega \left( W(\phi)+ U(\nabla \bfu, \phi) + \frac{1}{2} \kappa_\phi |\nabla \phi|^2 \right) \dm V
\end{equation}
where $\kappa_\phi$ is the phase-field gradient energy coefficient. The chemical free energy $W(\phi)$ is defined as $W(\phi)=(1-h(\phi)) f_\beta+h(\phi) f_\gamma+\Delta f G(\phi)$, where $f_\beta$ and $f_\gamma$ are the free energies of the $\beta$-Ti and $\gamma$-ZrC phases respectively, $\Delta f $ is the energy barrier height between the phases, $G(\phi)=16 \phi^2(1-\phi)^2$ is a double-well potential, and $h(\phi)=\phi^2(3-2 \phi)$ is a smooth interpolation function as defined in \cite{kovacevic_interfacial_2020}. The elastic energy density is $U(\nabla \bfu, \phi)=\frac{1}{2}\left(\varepsilon-\varepsilon_{eg}\right): \bfC:\left(\varepsilon-\varepsilon_{eg}\right)$, where the eigenstrain $\varepsilon_{eg}=[\phi(\bfx, t)-\phi(\bfx, 0)] \varepsilon^{\beta \rightarrow \gamma}$ accounts for phase transformation strains between $\beta$-Ti and $\gamma$-ZrC. The stiffness tensor $\bfC=\bfC_{\beta}+h(\phi)\left(\bfC_\gamma-\bfC_\beta\right)$ is interpolated between the stiffness of the individual phases, $\bfC_\gamma$ and $\bfC_\beta$.

The governing equations for the quasi-static mechanical equilibrium and phase-field evolution are:
\begin{align}
    \divergence \bfsigma &= 0 \\
    \frac{\partial \phi}{\partial t} &= (1-g(\phi))\left(-\frac{\partial W}{\partial \phi}-\frac{\partial U}{\partial \phi}+\kappa_\phi \nabla^2 \phi\right)
\end{align}
where the stress $\bfsigma = \partial U / \partial \varepsilon$. The phase-field mobility $M_\phi$ is taken as unity for the simulations presented, effectively scaling time or incorporating mobility into other parameters. The term $g(\phi)$ incorporates the geometric criteria discussed previously and can be temperature-dependent to control bonding.

\subsection{Weak Formulation}
The coupled system is solved using a mixed finite element approach. The weak forms for the displacement field $\bfu$, phase field $\phi$, and auxiliary gradient field $\bfpsi = \nabla \phi$ are:

Find $(\bfu_h, \phi_h, \bfpsi_h) \in V_h^\bfu \times V_h^\phi \times V_h^{\bfpsi}$ such that for all test functions $(\hat{\bfu}_h, \hat{\phi}_h, \hat{\bfpsi}_h) \in V_h^\bfu \times V_h^\phi \times V_h^{\bfpsi}$:
\begin{align}
    \text{Mechanical equilibrium:} \quad &\int_\Omega \bfsigma (\bfu_h, \phi_h) : \nabla \hat{\bfu}_h \, \dm V = \int_{\partial \Omega} \bft \cdot \hat{\bfu}_h \, dS \\
    \text{Auxiliary gradient field:} \quad &\int_\Omega \bfpsi_h \cdot \hat{\bfpsi}_h \, \dm V - \int_\Omega \nabla \phi_h \cdot \hat{\bfpsi}_h \, \dm V = 0 \label{eq:weak_form_psi_ti_zrc} \\
    \text{Phase-field evolution:} \quad &\int_\Omega \frac{\phi_h^{n+1}-\phi_h^n}{\Delta t} \hat{\phi}_h \, \dm V 
    + \int_\Omega \kappa_\phi \,\nabla \phi_h^{n+1}\!\cdot\nabla\!\big(\hat{\phi}_h\, m_h^n\big) \, \dm V
    + \int_\Omega m_h^n \left(\frac{\partial W}{\partial \phi}+\frac{\partial U}{\partial \phi}\right)_{\!n+\theta} \hat{\phi}_h \, \dm V = 0
    \label{eq:weak_form_phi_ti_zrc}
\end{align}
The auxiliary field $\bfpsi_h$ is updated at each time step via the weak constraint~\eqref{eq:weak_form_psi_ti_zrc}, implemented as an $L^2$-projection of $\nabla \phi_h$ onto $V_h^{\bfpsi}$ rather than as a simultaneous unknown in the Newton solve. 
The subscripts $n$ and $n+1$ denote time levels, and $n+\theta$ indicates a semi-implicit evaluation (e.g., $\theta=1$ for Backward Euler for the driving force terms).
The lagged mobility factor is defined as
\(m_h^n = (1-g_h^n)\), so that the geometric hindering enters both the local driving force and the diffusive flux through the term \(\nabla(\hat{\phi}_h m_h^n)\). The resulting scheme corresponds to a staggered mixed formulation in which $(\bfu_h,\phi_h)$ are solved implicitly and $\bfpsi_h$ is updated consistently after each phase-field solve.

\subsection{Simulation Parameters and Results}
\label{subsec:ti_zrc_sim_results}

To explore the model's behavior in the Ti-ZrC system, particularly the role of initial interlayer thickness and the influence of parameters, we simulated two primary initial Ti interlayer thicknesses: approximately $200\,\mu\text{m}$ and $50\,\mu\text{m}$. For these simulations, baseline chemical free energy parameters were set as $f_\beta=0.05$, $f_\gamma=0$, and $\Delta f=0.1$, with a phase-field gradient coefficient $\kappa_\phi=0.1$. The parameter $\alpha$ within $g(\phi)$ was set to $0.6$. Simulations were run until the interface evolution effectively ceased due to the activation of the coalescence term.

\subsubsection{Case 1: $200\,\mu\text{m}$ Interlayer}
\begin{figure}[htb!]
    \centering
    \subfloat[Initial configuration ($200\,\mu\text{m}$ interlayer).]{
        \includegraphics[width=0.45\textwidth]{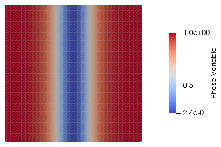}
        \label{fig:TiZrC200_a_results}
    }
    \hfill
    \subfloat[Final, arrested configuration.]{
        \includegraphics[width=0.45\textwidth]{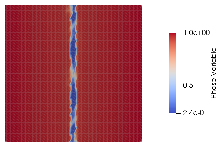}
        \label{fig:TiZrC200_b_results}
    }
    \caption{Comparison of the initial and final states for the $200\,\mu\text{m}$ Ti interlayer. Model parameters: $\kappa_\phi=0.1$, $f_\beta=0.05$, $\Delta f=0.1$, and $\varepsilon^{\beta\to\gamma}=0$.}
    \label{fig:TiZrC200_results}
\end{figure}

Figure~\ref{fig:TiZrC200_results} shows the initial and final states for the $200\,\mu\text{m}$ interlayer. The results show that $g(\phi)$ is activated when the interlayer ($\phi \approx 0$) shrinks below a certain thickness, thereby preventing complete homogenization and stabilizing a residual Ti layer.

\begin{figure}[htb!]
    \centering
    \subfloat[Strain, for $\varepsilon^{\beta \rightarrow \gamma} = 0$]{
        \includegraphics[width=0.3\textwidth]{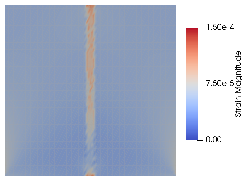}
    }
    \hfill
    \subfloat[Strain, for $\varepsilon^{\beta \rightarrow \gamma} = 10^{-4}$]{
        \includegraphics[width=0.3\textwidth]{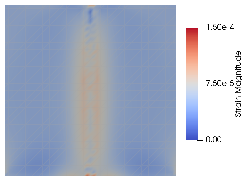}
    }
    \hfill
    \subfloat[Strain, for $\varepsilon^{\beta \rightarrow \gamma} = 10^{-3}$]{
        \includegraphics[width=0.3\textwidth]{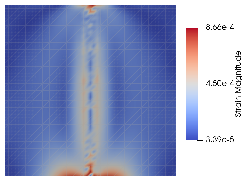}
    }
    \\
    \subfloat[Stress, for $\varepsilon^{\beta \rightarrow \gamma} = 0$]{
        \includegraphics[width=0.3\textwidth]{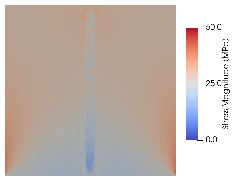}
    }
    \hfill
    \subfloat[Stress, for $\varepsilon^{\beta \rightarrow \gamma} = 10^{-4}$]{
        \includegraphics[width=0.3\textwidth]{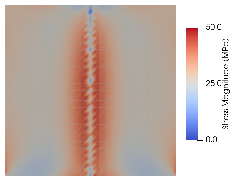}
    }
    \hfill
    \subfloat[Stress, for $\varepsilon^{\beta \rightarrow \gamma} = 10^{-3}$]{
        \includegraphics[width=0.3\textwidth]{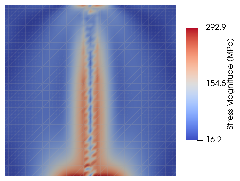}
    }
    \caption{Comparison of the strain (top row) and stress (bottom row) for various eigenstrain values in the $200\,\mu\text{m}$ interlayer case. Increasing eigenstrain leads to higher stresses and alters the strain localization.}
    \label{fig:StressStrainCompare_results}
\end{figure}

\begin{figure}[htb!]
    \centering
    \subfloat[$\phi$ with $\varepsilon^{\beta \rightarrow \gamma} = 0$]{
        \includegraphics[width=0.3\textwidth]{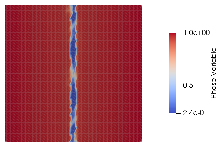}
    }
    \hfill
    \subfloat[$\phi$ with $\varepsilon^{\beta \rightarrow \gamma} = 10^{-4}$]{
        \includegraphics[width=0.3\textwidth]{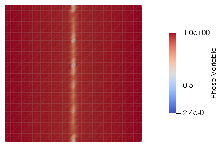}
    }
    \hfill
    \subfloat[$\phi$ with $\varepsilon^{\beta \rightarrow \gamma} = 10^{-3}$]{
        \includegraphics[width=0.3\textwidth]{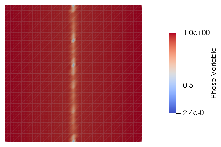}
    }
    \caption{Comparison of final state for various eigenstrain values. As the eigenstrain increases, the elastic energy modifies the pinned interface shape and location, influencing the final interlayer profile.}
    \label{fig:EigenstrainCompare_results}
\end{figure}

\paragraph{Effect of Eigenstrain.}
The influence of transformation-induced strain was examined by varying the eigenstrain $\varepsilon^{\beta \rightarrow \gamma}$ associated with the $\beta$-Ti to $\gamma$-ZrC transformation. Simulations were performed for $\varepsilon^{\beta \rightarrow \gamma} = 0$, $10^{-4}$, and $10^{-3}$. Figure~\ref{fig:StressStrainCompare_results} illustrates the corresponding stress and strain fields, revealing that increasing eigenstrain leads to more significant mechanical effects and stress accumulation. 
For these parameters, the final phase-field distributions (Figure~\ref{fig:EigenstrainCompare_results}) remain qualitatively similar between $\varepsilon^{\beta \rightarrow \gamma} = 10^{-4}$ and $10^{-3}$. This suggests that while the mechanical energy contribution scales with the eigenstrain, it only begins to locally perturb the pinned interface geometries without altering the global phase topology. This highlights a regime where chemical driving forces dominate the macroscopic morphological evolution, while the eigenstrain governs the magnitude of local internal stresses.


\paragraph{Effect of Chemical Free Energy Parameters $f_\beta$ and $\Delta f$.}

\begin{figure}[htb!]
    \centering
    \subfloat[Case (a): $f_\beta=0.05, \Delta f=0.1$]{
        \includegraphics[width=0.3\textwidth]{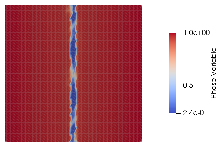}
        \label{fig:compareFbDf_a_results}
    }
    \hfill
    \subfloat[Case (b): $f_\beta=0.05, \Delta f=0.5$]{
        \includegraphics[width=0.3\textwidth]{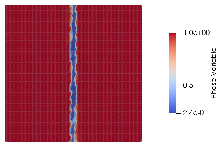}
        \label{fig:compareFbDf_b_results}
    }
    \hfill
    \subfloat[Case (c): $f_\beta=0.01, \Delta f=0.1$]{
        \includegraphics[width=0.3\textwidth]{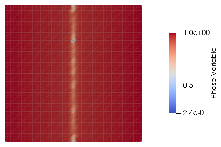}
        \label{fig:compareFbDf_c_results}
    }
    \caption{Final state for three different combinations of $f_\beta$ and $\Delta f$, with $f_\gamma=0$, $\kappa_\phi=0.1$, and $\varepsilon^{\beta\to\gamma}=0$. (a) $\{f_\beta=0.05, \Delta f=0.1\}$; (b) $\{f_\beta=0.05, \Delta f=0.5\}$; (c) $\{f_\beta=0.01, \Delta f=0.1\}$.}
    \label{fig:compareFbetaDeltaF_results}
\end{figure}

The interplay between the bulk-phase free energy of $\beta$-Ti ($f_\beta$) and the double-well potential amplitude ($\Delta f$) was examined, keeping $f_\gamma=0$ and $\kappa_\phi=0.1$. Three combinations were considered: (a) $f_\beta = 0.05, \Delta f=0.1$; (b) $f_\beta = 0.05, \Delta f=0.5$; (c) $f_\beta = 0.01, \Delta f=0.1$. Figure~\ref{fig:compareFbetaDeltaF_results} shows the final $\phi$ fields. 
Comparing Cases (a) and (b) demonstrates that increasing the barrier height $\Delta f$ while holding $f_\beta$ constant leads to a distinct and slightly wider transition zone. 
The higher energetic barrier in Case (b) results in a more pronounced pinning effect, as the system must overcome a larger local penalty to advance the transformation. 
Conversely, a comparison between Cases (a) and (c) shows that lowering $f_\beta$ (increasing Ti phase stability) leads to a significantly thinner interfacial region with a higher minimum phase field value. 
These results highlight that while $f_\beta$ determines the overall thermodynamic stability, $\Delta f$ controls the local sharpness and resistance to completion of the phase transition.

\subsubsection{Case 2: $50\,\mu\text{m}$ Interlayer and Tunability via coalescence Parameters}
\begin{figure}[htb!]
    \centering
    \subfloat[Initial state ($50\,\mu\text{m}$ interlayer).]{
        \includegraphics[width=0.45\textwidth]{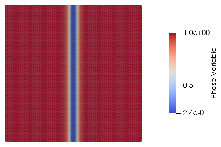}
        \label{fig:ZrC_50micron_a_results}
    }
    \hfill
    \subfloat[Final stabilized interface.]{
        \includegraphics[width=0.45\textwidth]{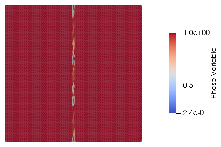}
        \label{fig:ZrC_50micron_b_results}
    }
    \caption{Evolution of the phase field for the $50\,\mu\text{m}$ Ti interlayer, with $\kappa_\phi=0.1$, $f_\beta=0.05$, $\Delta f=0.1$, and $\varepsilon^{\beta \rightarrow \gamma}=10^{-4}$. (a) Initial configuration. (b) Final arrested state resulting in a controlled gap.}
    \label{fig:ZrC_50micron_results}
\end{figure}

Simulations involving a thinner initial Ti interlayer of approximately $50\,\mu\text{m}$ were conducted using baseline parameters $\Delta f=0.1$, $f_\beta=0.05$, and an eigenstrain of $\varepsilon^{\beta \rightarrow \gamma}=10^{-4}$. The results, shown in Figure~\ref{fig:ZrC_50micron_results},  demonstrate the ability to control the final interlayer width. Critically, by adjusting the threshold parameters $\alpha, \beta,$ and $\gamma$, the interface evolution can be effectively arrested at a predetermined final gap. This capability is particularly valuable for calibrating the model against experimental observations of specific interlayer thicknesses or for designing processes where a specific residual interlayer is desired, thereby implicitly capturing complex thermodynamic or kinetic barriers to complete coalescence.

\vspace{2mm}

\section{Discussion}
\label{sec:discussion}

The kinetic phase-field model presented in this work introduces an approach to controlling interface coalescence, particularly relevant to simulating processes like diffusion bonding.
Our approach is to systematically modulate the interface kinetics through a coalescence kinetic function.
This kinetic function depends on nonlocal higher-gradient geometric variables of the phase field (such as its gradient, Laplacian, and Hessian), following similar strategies of using nonlocal gradient terms to describe complex behavior \cite{deshmukh2022multiband,dayal2017leading}.
This allows the arrest of interface motion in regions where specific geometric criteria -- indicative of desired interface stability or a barrier to coalescence -- are met. These criteria can be linked to macroscopic experimental parameters like temperature or critical interlayer thickness, thereby allowing the model to implicitly reflect the governing thermodynamic conditions for joining or non-joining.
This enables the prediction of quantities such as the final interface thickness and morphology based on energetics and kinetics, without explicitly tracking solute concentration fields or their gradients.

This approach has distinct features  when compared to phase-field models that explicitly couple phase evolution with solute transport via, for example, a Cahn-Hilliard framework, such as the model for Ti-ZrC by Kovacevic et al. \cite{kovacevic_interfacial_2020}. 
Their work relies on carbon concentration as a primary field and its influence on interfacial gradient energy as a driving force for phenomena like critical interlayer thickness, necessitating the solution of a fourth-order Cahn-Hilliard equation alongside the phase-field and mechanical equations. 
Our formulation, by omitting the explicit solute field, reduces the number of coupled partial differential equations to be solved. While it does not resolve the detailed solute profiles, it aims to capture the macroscopic structural outcome of phase transformations and interface interactions. This can be computationally advantageous, particularly for scenarios where the detailed concentration pathway is of secondary interest compared to the final interface stability or morphology under given conditions. 

The interface identification proposed in this work also offers an approach to simulating systems where interface behavior is strongly dictated by nonlocal geometry and overarching thermodynamic conditions, potentially applicable to a diverse range of joining or other phase transformation problems in intermetallics, functionally-graded materials, and CMCs.
For example, the model could be applied to simulate grain boundary pinning, where the motion is arrested upon geometric contact with a second-phase particle, or arrested crack healing, where a crack interface is preserved until specific thermodynamic conditions are met.

\section*{}

\paragraph*{Software and Data Availability.}

The code developed for this work and the associated data are available at \\ \url{https://github.com/maryam-khd/fenics-diffusion-bonding}

\paragraph*{Acknowledgments.}

We thank ARO (MURI W911NF-24-2-0184) and the Pennsylvania Infrastructure Technology Alliance for financial support; and NSF ACCESS (MCH240078) for computing resources provided by the Pittsburgh Supercomputing Center.

\appendix
\section{Selection of Model Parameters}
\label{app:parameters}

This appendix details the rationale behind the selection of key phase-field model parameters used in the numerical simulations. 

\begin{enumerate}
    \item \textbf{Gradient Energy Coefficient ($\kappa_\phi$):}
    The parameter $\kappa_\phi$ determines the energetic penalty for spatial gradients in $\phi$, thereby influencing the diffuse interface thickness. It is essential that this interface spans several grid cells to be numerically well-resolved. If the physical interface is much thinner than the mesh resolution ($\Delta x$), $\kappa_\phi$ is often chosen based on $\Delta x$. In this study, $\kappa_\phi$ was selected such that the nominal interface width was proportional to $5\Delta x$ to $10\Delta x$. For instance, with $\kappa_\phi = 0.1$ and $\Delta x = 0.02$, in dimensionless quantities. This ensures that the diffuse interface is adequately captured by the discrete mesh without being overly diffuse or numerically unstable.

    \item \textbf{Reference Chemical Energy ($f_\gamma$):}
    One of the bulk phases is typically chosen as an energetic reference. In this work, the $\gamma$-phase (e.g., ZrC in the Ti-ZrC system, represented by $\phi \approx 1$) was set to have zero baseline chemical energy: $f_\gamma = 0$.
    This simplifies the interpretation of $f_\beta$ as the relative energy difference.

    \item \textbf{Relative Chemical Energy ($f_\beta$):}
    The chemical energy of the $\beta$-phase (e.g., Ti, represented by $\phi \approx 0$) relative to the $\gamma$-phase was selected in relation to $\kappa_\phi$ and $\Delta f$. For the specific parameters presented ($f_\beta=0.05$, $\kappa_\phi=0.1$, $\Delta f=0.1$), this sets a moderate energy penalty for the $\beta$-phase.

    \item \textbf{Double-Well Potential Height ($\Delta f$):}
    The parameter $\Delta f$ scales the height of the energy barrier in the double-well potential $G(\phi)$, influencing the energetic cost of intermediate phase-field values ($0 < \phi < 1$). It is typically chosen to be of a similar order of magnitude as the interfacial energy density per unit volume (related to $\kappa_\phi/\ell^2$) or the bulk energy differences. Setting $\Delta f$ appropriately with respect to $\kappa_\phi$ (e.g., $\Delta f = \kappa_\phi = 0.1$ as used in simulations) ensures a clear energetic distinction between the bulk phases and the interfacial region, promoting phase separation.
\end{enumerate}

\newcommand{\etalchar}[1]{$^{#1}$}

\end{document}